\documentclass[conference,a4paper,10pt]{IEEEtran}
\usepackage[margin = 1.5cm]{geometry}

\usepackage[utf8]{inputenc} 
\usepackage[T1]{fontenc}
\usepackage{url}
\usepackage{ifthen}
\usepackage[noadjust]{cite}
\usepackage[cmex10]{amsmath}
\usepackage{multirow}
\usepackage{cite}
\usepackage{amsmath,amssymb,amsfonts}
\usepackage{algorithm}
\usepackage{algpseudocode}
\usepackage{graphicx}
\usepackage{makecell}
\usepackage{textcomp}
\usepackage{xcolor}
\usepackage{caption}
\usepackage{mathtools}
\usepackage{comment}
\usepackage{amsthm}
\usepackage{csquotes}
\usepackage{enumitem}
\usepackage{romannum}
\usepackage{stfloats}
\newtheorem{thm}{Theorem}
\newtheorem{lem}{Lemma}

\newtheorem{corollary}{Corollary}
\newtheorem{rem}{Remark}

\newtheorem{defn}{Definition}

\newtheorem{exmp}{Example}

\begin{document}
\title{Extended Placement Delivery Arrays for Multi-Antenna Coded Caching Scheme} 

\author{\IEEEauthorblockN{K. K. Krishnan Namboodiri, Elizabath Peter and B. Sundar Rajan}
	\IEEEauthorblockA{\textit{Department of Electrical Communication Engineering} \\
		\textit{Indian Institute of Science}, Bengaluru, India \\
		Email: \{krishnank,elizabathp, bsrajan\}@iisc.ac.in}
}

\maketitle

\begin{abstract}
The multi-antenna coded caching problem, where the server having $L$ transmit antennas communicating to $K$ users through a wireless broadcast link, is addressed. In the problem setting, the server has a library of $N$ files, and each user is equipped with a dedicated cache of capacity $M$. The idea of extended placement delivery array (EPDA), an array which consists of a special symbol $\star$ and integers in a set $\{1,2,\dots,S\}$, is proposed to obtain a novel solution for the aforementioned multi-antenna coded caching problem. From a $(K,L,F,Z,S)$ EPDA, a multi-antenna coded caching scheme with $K$ users, and the server with $L$ transmit antennas, can be obtained in which the normalized memory $\frac{M}{N}=\frac{Z}{F}$, and the delivery time $T=\frac{S}{F}$. The placement delivery array (for single-antenna coded caching scheme) is a special class of EPDAs with $L=1$. For the multi-antenna coded caching schemes constructed from EPDAs, it is shown that the maximum possible Degree of Freedom (DoF) that can be achieved is $t+L$, where $t=\frac{KM}{N}$ is an integer. Furthermore, two constructions of EPDAs are proposed: a) $ K=t+L$, and b) $K=nt+(n-1)L, \hspace{0.1cm}L\geq t$, where $n\geq 2$ is an integer. In the resulting multi-antenna schemes from those EPDAs achieve the full DoF, while requiring a subpacketization number $\frac{K}{\text{gcd}(K,t,L)}$. This subpacketization number is less than that required by previously known schemes in the literature.
\end{abstract}

\IEEEpeerreviewmaketitle
\section{Introduction}
\label{intro}
Coded caching, introduced in \cite{MaN}, is a promising technique to reduce the peak hour network traffic in content delivery networks by exploiting the caches at the user end. This technique involves two phases, namely the placement phase and the delivery phase. During the placement phase, the central server (having a library of $N$ files) populates the caches at the user end. There are $K$ users, each having a cache of size $M$ units, where $0\leq M\leq N$. In the delivery phase, each user requests a single file from the server library.  After knowing those requests, the server transmits a message created by encoding across the requested file contents. The setting in \cite{MaN} considered an error-free broadcast link between the server (with a single transmit antenna) and the users. The normalized capacity of the link was assumed to be one file per unit of time. The scheme in \cite{MaN} can meet all the user demands in a normalized delivery time of $T=\frac{K-t}{t+1}$, where $t=\frac{KM}{N}$ is an integer. The term $t+1$ in the denominator shows the multiplicative reduction obtained in the delivery time by the scheme in \cite{MaN} compared to the conventional uncoded caching (where the users will be served one after the other- coding across the demanded files is not employed). The term $t+1$ is termed as the global caching gain or the degree of freedom (DoF) achieved by the scheme. The DoF achieved by a scheme is defined as the number of users simultaneously served in unit time. In the further discussions, we refer to the scheme proposed in \cite{MaN} as the MaN scheme. The MaN scheme achieves the DoF $\frac{KM}{N}+1$ by splitting the finite-length files into $\binom{K}{t}$ subfiles. The exponentially increasing subpacketization issue was addressed in \cite{YCTC} by constructing coded caching schemes from placement delivery arrays (PDA). But the reduction in subpacketization number was achieved at the cost of the DoF. Furthermore, in \cite{YCTC}, the authors showed that the MaN scheme could also be obtained from a class of PDAs (we refer to that class of PDAs as the MaN PDA).

The coded caching problem with multiple transmit antennas (with the server) were explored in \cite{SMK,SCK}, and proposed a scheme that achieves a DoF of $t+L$, where $t=\frac{KM}{N}$ is an integer and $L$ is the number of transmit antennas. Under the assumption of under uncoded cache placement and one-shot data delivery $t+L$ (where $t+L\leq K$) is proven to be the optimal DoF  \cite{LaE1}. For the schemes in \cite{SMK,SCK}, to achieve this theoretical DoF, a file has to be split into,
$$\binom{K}{t}\binom{K-t-1}{L-1}$$  
subfiles. Even though those schemes enabled to achieve the optimal DoF, the subpacketization requirement was far more than the MaN scheme. The multi-antenna coded caching scheme proposed in \cite{LaE} tackled the subpacketization bottleneck without paying in the DoF. The presented scheme required only a subpacketization number of $\binom{K/L}{t/L}$ while achieving the full DoF of $t+L$. This subpacketization number is approximately the $L^{th}$ root of the subpacketization number required for the MaN scheme. But, the scheme is valid only when $L|K$ ($L$ divides $K$) and $L|t$. The schemes in \cite{STSK,STS,SPSET} address the multi-antenna coded caching problem under the subpacketization constraint. In \cite{STSK}, the trade-off between the delivery time and the subpacketization number is investigated for the case with $K=t+L$. The scheme proposed in \cite{STS} achieves the DoF $t+L$ with a linear subpacketization number of $K(t+L)$, when $L\geq t$. In \cite{SPSET}, the authors proposed a scheme that achieves the DoF $t+\alpha$, where $t\leq \alpha \leq L$. The subpacketization requirement is $\frac{K(t+L)}{\gamma^2}$, where $\gamma = \text{gcd}(K,t,L)$ , which is even less compared to the subpacketization requirement in \cite{STS}. Other interesting works in multi-antenna coded caching literature include \cite{TSKK,TSKK2} that consider optimized precoder design, \cite{ZAG,LZSE} that reduce the complexity by limiting the number of messages received by the users in every time slot.

\subsection{Contributions}
In this work, we study the coded caching problem where the server has $L$ transmit antennas. We provide an alternate solution for the multi-antenna coded caching problem by designing a special array termed as extended placement delivery array (EPDA). The technical contributions of this paper are summarized:
\begin{itemize}
	\item The idea of EPDA is proposed (Section \ref{ata}: Definition \ref{defn:ata}). The placement delivery array (PDA) proposed in \cite{YCTC} is a special class of EPDAs with $L=1$ (Section \ref{ata}: Remark \ref{rem:ata}).
	\item A novel solution for the multi-antenna coded caching problem via EPDAs is provided. It is shown that for every EPDA there exists a corresponding multi-antenna coded caching scheme (Section \ref{ata}: Theorem \ref{thm:ata}).
	\item For the multi-antenna coded caching schemes constructed from regular EPDAs (an EPDA is said to be regular if all the integers in the array appear exactly the same number of times), it is shown that the maximum possible DoF that can be achieved is $t+L$ (Section \ref{ata}: Lemma \ref{lem:reg}).
	\item A class of EPDAs corresponding to the coded caching scheme proposed in \cite{LaE} is identified (Section \ref{constructions}: Remark \ref{rem:LaE}).
	\item Two constructions of EPDAs are proposed. The resulting multi-antenna coded caching schemes are applicable
	\begin{enumerate}
		\item when $K=t+L$, (Section \ref{constructions}: Construction \Romannum{1})
		\item when $K=nt+(n-1)L, \hspace{0.1cm}L\geq t$, where $n\geq 2$ is a positive integer (Section \ref{constructions}: Construction \Romannum{2}).
	\end{enumerate} 
	\item The multi-antenna coded caching schemes resulting from Construction \Romannum{1} and Construction \Romannum{2} achieve full DoF $t+L$, while requiring a subpacketization number $\frac{K}{\text{gcd}(K,t,L)}$ (Section \ref{constructions}: Theorem \ref{thm:constA}, Section \ref{constructions}: Theorem \ref{thm:constB}). The subpacketization number for those schemes is less than that required by previously known schemes in the literature (Section \ref{comparison}: Table \ref{table:constA}, Section \ref{comparison}: Table \ref{table:constB}).
\end{itemize}

Our primary focus is on the construction of multi-antenna coded caching schemes from EPDAs. We are mainly interested in DoF analysis (i.e., at higher SNR) rather than the design of sophisticated beamformers. 

\subsection{Notations}
For a positive integer $n$, $[n]$ denotes the set $ \left\{1,2,\hdots,n\right\}$. For two positive integers $a,b$ such that $a\leq b$, $[a:b] = \{a,a+1,\hdots,b\}$. For integers $a,b\leq K$, 
\begin{equation*}
[a:b]_K =
\begin{cases}
\left\{a,a+1,\hdots,b\right\} & \text{if } a\leq b.\\
\left\{a,a+1,\hdots,K,1,\hdots,b\right\} & \text{if } a>b.
\end{cases}   
\end{equation*}
For any two integers, $i$ and $K$, 
\begin{equation*}
<i>_K =
\begin{cases}
i\text{ }(mod\text{ }K) & \text{if $i$ $(mod$ $K) \neq0$. }\\
K & \text{if $i$ $(mod$ $K) =0$.}
\end{cases}   
\end{equation*}
For two positive integers $a$ and $b$, $\text{gcd}(a,b)$ represents the greatest common divisor of $a$ and $b$, and $a|b$ means that $a$ divides $b$ (i.e., $b=pa$ for some integer $p$). Similarly, $\text{gcd}(a,b,c)$ represents the greatest common divisor of three positive integers $a,b$ and $c$. For two vectors $\mathbf{u}$ and $\mathbf{v}$, $\mathbf{u}\perp \mathbf{v}$ means that $\mathbf{v}^T\mathbf{u}=0$, and $\mathbf{u}\not\perp \mathbf{v}$ means that $\mathbf{v}^T\mathbf{u}\neq0$. All the vectors are assumed to be column vectors by default. Finally, the symbol $\mathbb{C}$ represents a complex number. 

\section{System Model and Problem Formulation}
\label{system}
\begin{figure}[t]
	\begin{center}
		\captionsetup{justification = centering}
		\includegraphics[width = 0.75\columnwidth]{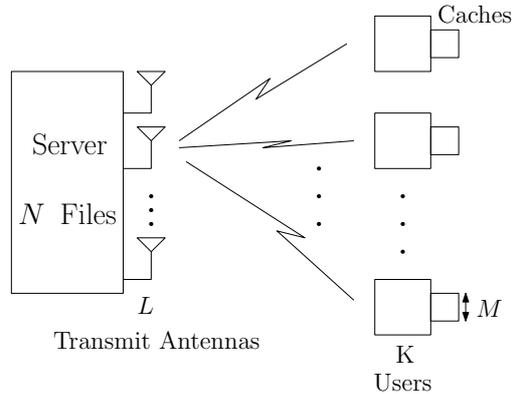}
		\caption{$(K,L,M,N)$ Multi-antenna Coded Caching System }
		\label{MACC}
	\end{center}
\end{figure}

The system model consists of a central server having a library of $N$ files, $W_{[1:N]}\triangleq \{W_n:n\in [N]\}$ each of size 1 unit. We consider a multiple-input, single-output (MISO) broadcast channel in which the server with $L$ transmit antennas communicates to $K$ users, each having a single receive antenna. The wireless shared link is assumed to be of capacity 1 file per unit of time. Furthermore, each node in the system (the server and $K$ users) has perfect channel state information (CSI). Each user is equipped with a dedicated cache of capacity $M$ units, where $0\leq M\leq N$ (see Fig. \ref{MACC}). The system operates in two phases: the placement phase and the delivery phase. In the placement phase, the server stores some of the file contents in the caches of the users without knowing the future demands. In general, the placement can be coded or uncoded. In this work, we concentrate only on schemes with uncoded placement. The contents stored in cache $k$ is denoted as $\mathcal{Z}_k$. In the delivery phase, each user requests a single file from the server. Let $\mathbf{d} = (d_1,d_2,\dots,d_K)$ be the demand vector, i.e., the $k^{th}$ user requests for the file $W_{d_k}$, for every $k\in [K]$. After knowing the demand vector, the server makes transmission vectors $\{\textbf{x}(\tau)\}_{\tau=1}^T$. That is, the server transmission is for $T$ time slots. During the time slot $\tau$, the server transmits $\textbf{x}(\tau)$, where $\textbf{x}(\tau)\in \mathbb{C}^L$. At time slot $\tau$, the $k^{th}$ user receives,
\begin{equation}
y_k(\tau) = \mathbf{h}_k^T\textbf{x}(\tau)+w_k(\tau)
\end{equation}
where $\mathbf{h}_k^T\in\mathbb{C}^L$ is the channel vector and $w_k(\tau)\sim \mathbb{C}\mathcal{N} (0,1)$ is the additive noise (complex normal distributed with zero mean and unit variance) observed at user $k$ at time slot $\tau$. Define the channel matrix $\mathbf{H} := [\mathbf{h}_1,\mathbf{h}_2,\dots,\mathbf{h}_K]$. We assume that the received signal to noise ratio (SNR) is high as in \cite{SMK,LaE,STSK,STS}, and neglect the additive noise component during the analysis. The correctness of the scheme implies that, using the local cache content $\mathcal{Z}_k$ and the received coded files $y_k(1),y_k(2),\dots,y_k(T)$, user $k$ should be able to decode the demanded file $W_{d_k}$. The coded caching system under the aforementioned setting is called the $(K,L,M,N)$ multi-antenna coded caching system. 

The number of time slots $T$ taken by the server to meet the user demands is termed the delivery time. It is already proven in the literature \cite{LaE1} that under the assumption of uncoded placement and one-shot delivery, the optimal delivery time
\begin{equation}
T^* = \frac{K-t}{t+L}
\end{equation} 
where, $t=\frac{KM}{N}$ is an integer. 
For any other value of $M$ (when $\frac{KM}{N}$ is not an integer), the delivery time is the lower convex envelope of adjacent corner points (connect the delivery times at the adjacent integer $t$ values with a straight line). The term $1-\frac{t}{K}=1-\frac{M}{N}$ denotes the local caching gain, which is achieved simply from the caching (in the placement phase). The extra multiplicative reduction factor $t+L$ is termed as the Degree of Freedom (DoF) achieved. The ultimate goal of the coded caching problem is to design the placement and the delivery phases jointly such that the DoF is maximized (maximizing the DoF is the same as minimizing the delivery time). 

While designing coded caching schemes, one other important parameter to be considered is the subpacketization number. The subpacketization number is defined as the number of subfiles into which a file is divided during the coded caching scheme (considering both placement and delivery phases). It is always desired to have a low subpacketization number.

\section{Preliminaries}
\label{prelims}

In this section, we briefly review PDAs and see how the MaN scheme can be obtained from a special class of PDAs (\cite{YCTC}).

\subsection{Placement Delivery Array (PDA)}
\begin{defn}
	\label{def:pda}
	(\cite{YCTC}) For positive integers $K, F, Z$ and $S$, an $F \times K$ array $\mathbf{P}=[p_{j,k}]$, $j \in [F]$ and $k \in [K]$, composed of a specific symbol $\star$ and $S$ positive integers $1,2,\ldots, S$, is called a $(K,F,Z,S)$ placement delivery array (PDA) if it satisfies the following three conditions: \\
	\textit{D1}. The symbol $\star$ appears $Z$ times in each column.\\
	\textit{D2}. Each integer occurs at least once in the array.\\
	\textit{D3}. For any two distinct entries $p_{j_1,k_1}$ and $p_{j_2,k_2}$, $p_{j_1,k_1}=p_{j_2,k_2}=s$ is an integer only if
	\begin{enumerate}[label=(\alph*)]
		
		\item $j_1 \neq j_2$, $k_1 \neq k_2$, i.e., they lie in distinct rows and distinct columns, and
		\item $p_{j_1,k_2}=p_{j_2,k_1}=\star$, i.e., the corresponding $2\times2$ sub-array formed by rows $j_1, j_2$ and columns $k_1,k_2$ must be of the following form:\\
		\begin{center}
			
			$\begin{pmatrix}
			s & \star\\
			\star & s
			\end{pmatrix}$
			\hspace{0.3cm}or\hspace{0.3cm}
			$\begin{pmatrix}
			\star & s \\
			s & \star
			\end{pmatrix}$ 
			
		\end{center}

	\end{enumerate}
\end{defn}

Every $(K,F,Z,S)$ PDA corresponds to a coded caching scheme (single antenna, $L=1$) with parameters $K,M$ and $N$ as in Lemma~\ref{lemm:1}.

\begin{lem}
	(Theorem 1 in \cite{YCTC}) For a given $(K, F, Z, S)$ PDA $\mathbf{P}=[p_{j,k}]_{F \times K}$, a $(K,M,N)$ coded caching scheme can be obtained with subpacketization number $F$ and $\frac{M}{N}=\frac{Z}{F}$ using Algorithm 1. For any distinct demand vector $\mathbf{d}$, the demands of all the users are met with a delivery time, $T=\frac{S}{F}$.
	\label{lemm:1}
\end{lem}

\begin{algorithm}
	\renewcommand{\thealgorithm}{1}
	\caption{Coded caching scheme based on PDA \cite{YCTC}}
	\label{algpda}
	\begin{algorithmic}[1]
		\Procedure{Placement}{$\mathbf{P},W_{[1:N]}$}       
		\State Split each file $W_n$, $n \in [N]$ into $F$ subfiles: $W_n =\{W_{n,j}: j \in [F]\}$
		\For{\texttt{$k \in [K]$}}
		\State  $\mathcal{Z}_k$ $\leftarrow$ $\{W_{n,j}, \forall n \in [N]$: $p_{j,k}=\star, j \in [F]\}$
		\EndFor
		\EndProcedure
		
		\Procedure{Delivery}{$\mathbf{P},W_{[1:N]},\mathbf{d}$} 
		\For{\texttt{$s \in [S]$}}
		\State Server sends $\underset{\substack{p_{j,k}=s \\j\in [F],  \textrm{\hspace{0.05cm}} k\in[K]}}{\bigoplus}W_{d_k,j}$
		\EndFor    
		\EndProcedure
	\end{algorithmic}
\end{algorithm}

In a $(K,F,Z,S)$ PDA $\mathbf{P}$, the rows represent subfiles and the columns represent users. For any $k \in [K]$ if $p_{j,k}=\star$, then it implies that the user $k$ has access to the $j^{th}$ subfile of all the files. The contents placed in the $k^{th}$ user's cache is denoted by $\mathcal{Z}_k$ in Algorithm 1. If $p_{j,k}=s$ is an integer, then it means that the user $k$ does not have access to the $j^{th}$ subfile of any of the files. Condition $D1$ guarantees that all users have access to some $Z$ subfiles of all the files. According to the delivery procedure in Algorithm $1$, the server sends a linear combination of the requested subfiles indicated by the integer $s$ in the PDA. Therefore, condition $D2$ implies that the number of messages transmitted by the server is exactly $S$, and the delivery time is $\frac{S}{F}$. Condition $D3$ ensures the decodability. The placement and the delivery procedure based on PDA is given in Algorithm \ref{algpda}.

If all the integers are appearing exactly $g$ times in a PDA, then the PDA is said to be $g$-regular.
The following lemma shows that the MaN scheme can be obtained from a regular PDA.
\begin{lem}
(Theorem 2 in \cite{YCTC}) For a $(K,M,N)$ caching system with $\frac{M}{N} \in \{0, \frac{1}{K}, \frac{2}{K}, \ldots, 1 \}$, let $t=\frac{KM}{N}$, there exists a $t+1$ regular-$(K,F,Z,S)$ PDA with $F=\binom{K}{t}$, $Z=\binom{K-1}{t-1}$ and $S=\binom{K}{t+1}$.
\end{lem}

In the MaN scheme, choose $t \in [0:K]$ such that $F = \binom{K}{t}$. Each row of the PDA is indexed by sets $\mathcal{T} \subseteq [K]$ where $|\mathcal{T}|=t$. Each user stores subfiles $W_{n,\mathcal{T}}, \forall n \in [N]$ if $k \in \mathcal{T}$, thus $Z=\binom{K-1}{t-1}$. In the delivery phase of the MaN scheme, there are $\binom{K}{t+1}$ transmissions. Therefore, $\binom{K}{t+1}$ distinct integers are required to represent each of those transmissions resulting in $S = \binom{K}{t+1}$. To fill the PDA with these integers, a bijection $f$  is defined from the $t+1$ sized subsets of $\{1,2,\ldots,K\}$ to the set $\{1,2,\ldots,S\}$ such that 
\begin{equation}
p_{\mathcal{T},k} = 
\begin{cases}
f(\mathcal{T}\cup \{k\}), & \textrm{if \hspace{0.05cm}} k \notin \mathcal{T}.\\
\star, & \textrm{elsewhere}.
\end{cases}
\label{eq:pda_MN}
\end{equation}

From the above expression, it can be seen that each integer appears exactly $t+1$ times. That is, the regularity of the PDA is $t+1$.

\section{Extended Placement Delivery Array (EPDA)}
\label{ata}
In this section, we propose the idea of extended placement delivery array to obtain multi-antenna coded caching schemes. 
\begin{defn}
	\label{defn:ata}
	Let $K,L (\leq K), F,Z, S$ be positive integers. An array $\textbf{A}=[a_{j,k}]$, $j\in [F]$, $k\in [K]$ consisting of the symbol $\star$ and positive integers in $[S]$ is called a $(K,L,F,Z,S)$ extended placement delivery array (EPDA) if it satisfies the following conditions:\\
	C1. The symbol $\star$ appears $Z$ times in each column.\\
	C2. Every integer in the set $[S]$ occurs at least once in $\textbf{A}$.\\ 
	C3. No integer appears more than once in any column.\\
	C4. Consider the sub-array $\textbf{A}^{(s)}$ of $\textbf{A}$ obtained by deleting all the rows and columns of $\textbf{A}$ that do not contain the integer $s$. Then for any $s\in [S]$, no row of $\textbf{A}^{(s)}$ contains more than $L$ integers.
\end{defn}

 We present an example of an EPDA before discussing the construction of multi-antenna coded caching scheme from EPDAs. A $(K=3,L=2,F=3,Z=1,S=2)$ EPDA is given below:
\[
\textbf{A} = \begin{bmatrix}
\star & 1     & 1 \\
1     & \star & 2 \\
2     & 2     & \star 
\end{bmatrix}
\]
It is easy to verify that $\textbf{A}$ satisfies the conditions \textit{C1, C2} and \textit{C3}. To verify \textit{C4}, consider the sub-arrays $\textbf{A}^{(1)}$ and $\textbf{A}^{(2)}$. Since the integer 1 is not present in the third row of $\mathbf{A}$, the sub-array $\textbf{A}^{(1)}$ is obtained by deleting the third row of $\mathbf{A}$. Similarly, the sub-array $\textbf{A}^{(2)}$ is obtained by deleting the first row of $\mathbf{A}$.
\[
\textbf{A}^{(1)} = \begin{bmatrix}
\star & 1     & 1 \\
1     & \star & 2 
\end{bmatrix}, \hspace{0.5cm}
\textbf{A}^{(2)} = \begin{bmatrix}
1     & \star & 2 \\
2     & 2     & \star 
\end{bmatrix} 
\]
In both $\textbf{A}^{(1)}$ and $\textbf{A}^{(2)}$, all the rows contain two integers (which is equal to $L$). Therefore, $\textbf{A}$ satisfies the condition \textit{C4} as well. Note that, in the sub-array corresponding to the integer 1, integer 2 is also present. In general, a sub-array $\textbf{A}^{(s)}$ of a $(K,L,F,Z,S)$ EPDA may contain the integers in $[S]\backslash \{s\}$ as well.

From a $(K,L,F,Z,S)$ EPDA, $\textbf{A}=[a_{j,k}]$, $j\in[F]$ and $k\in [K]$, a multi-antenna coded caching scheme with the server having $L$ transmit antennas and the normalized cache size $\frac{M}{N}=\frac{Z}{F}$, can be obtained as follows:
\begin{enumerate}[label=\arabic*.]
	\item \textbf{Placement phase:} The server divides each file into $F$ subfiles of equal size. Thus, for every $n\in [N]$, we have, $W_n = \{W_{n,j}|j\in[F]\}$. The $k^{th}$ user's cache is populated as follows:
	\begin{equation}
	\label{ata:placement}
	\mathcal{Z}_k = \left\{W_{n,j}\hspace{0.1cm}|\hspace{0.1cm} a_{j,k}=\star, \hspace{0.15cm} \forall n\in [N]\right\}.
	\end{equation}
    For any $k \in [K]$ if $a_{j,k}=\star$, then it means that user $k$ has access to the $j^{th}$ subfile of all the files. There are $Z$ $\star$'s in the $k^{th}$ column of $\textbf{A}$ (by \textit{C1}). Since each subfile is of size $\frac{1}{F}$ units, the cache size of each user is $M = \frac{NZ}{F}$ units. Hence, $\frac{M}{N}= \frac{Z}{F}$.
	\item \textbf{Delivery phase:} Let $\mathbf{d}=(d_1,d_2,\dots,d_K)$ be the demand vector. Assume that the integer $s$ appears $g_s$ times in $\mathbf{A}$. Let $a_{j_1,k_1}=a_{j_2,k_2}=\dots=a_{j_{g_s},k_{g_s}}=s$. Then the server transmits $\mathbf{V}^s\cdot(W_{d_{k_1},j_1},W_{d_{k_2},j_2},\dots,W_{d_{k_{g_s}},j_{g_s}})^T$, where $\mathbf{V}^s=(\mathbf{v}^s_{1},\mathbf{v}^s_{2},\dots,\mathbf{v}^s_{{g_s}})$ is a precoding matrix of size $L\times g_s$. For every $i\in [g_s]$, define $\mathcal{B}_i:=\{\beta:  a_{j_i,\beta}\neq \star,\hspace{.1cm} \beta \in \{k_1,\dots,k_{i-1},k_{i+1},\dots,k_{g_s}\}\}$. Then the $i^{th}$ column of $\mathbf{V}^s$ is $\mathbf{v}^s_{i} \perp \mathbf{h}_\alpha$ for all $\alpha \in \mathcal{B}_i$, and $\mathbf{v}^s_{i} \not\perp \mathbf{h}_{k_i}$.
\end{enumerate}
 The placement and the delivery procedures are summarized in Algorithm \ref{alg1}.
\begin{algorithm}
	\caption{Multi-antenna coded caching scheme based on EPDA}
	\label{alg1}
	\begin{algorithmic}[1]
		\Procedure{Placement }{$\mathbf{A},W_{[1:N]}$}       
		\State Split each file $W_n,n \in [N]$ into $F$ subfiles: 
		$\textrm{\hspace{1cm}}W_n \leftarrow \{W_{n,j}: j \in [F]\}$
		\For{\texttt{$k \in [K]$}}
		\State  $\mathcal{Z}_k$ $\leftarrow$ $\{W_{n,j}, \forall n \in [N]$: $a_{j,k}=\star, j \in [F]\}$
		\EndFor
		\EndProcedure
		\Procedure{Delivery }{$\mathbf{A},W_{[1:N]},\mathbf{d},\mathbf{H}$} 
		\For{\texttt{$s \in [S]$}}
		\State $\textbf{A}^s \leftarrow [a_{j_i,k_i}], \textrm{\hspace{.4cm}}i\in[g_s]$ such that $\textrm{\hspace{2.1cm}}a_{j_1,k_1}=a_{j_2,k_2}=\dots=a_{j_{g_s},k_{g_s}}=s$
		\For{\texttt{$i \in [g_s]$}}
		\State $\mathcal{B}_i \leftarrow \{\beta:  a_{j_i,\beta}\neq \star,\textrm{\hspace{2.5cm}}$\hspace{5cm} $\textrm{\hspace{3cm}}\beta \in \{k_1,\dots,k_{i-1},k_{i+1},\dots,k_{g_s}\}\}$
		\State Design $\mathbf{v}^s_{i}$ such that $\mathbf{v}^s_{i} \perp \mathbf{h}_\alpha$ for all $\alpha \in \mathcal{B}_i$ and $\mathbf{v}^s_{i} \not\perp \mathbf{h}_{k_i}$
		\EndFor 
		\State $\mathbf{V}^s\leftarrow (\mathbf{v}^s_{1},\mathbf{v}^s_{2},\dots,\mathbf{v}^s_{{g_s}})$
		\State Transmit $\mathbf{V}^s\cdot(W_{d_{k_1},j_1},W_{d_{k_2},j_2},\dots,W_{d_{k_{g_s}},j_{g_s}})^T$
		\EndFor    
		\EndProcedure
	\end{algorithmic}
\end{algorithm}

For the above placement and delivery strategies, we have the following theorem.
\begin{thm}
	\label{thm:ata}
	Corresponding to any $(K,L,F,Z,S)$ EPDA, there exists a $(K,L,M,N)$ multi-antenna coded caching scheme with $\frac{M}{N}=\frac{Z}{F}$ and subpacketization number $F$. Furthermore, the server can meet any user demand $\mathbf{d}$ with a delivery time $T = \frac{S}{F}$.
\end{thm}
\begin{IEEEproof}
	The server stores $Z$ subfiles of every file in the $k^{th}$ user's cache  (Eq:\eqref{ata:placement}) during the placement phase. Therefore, the normalized memory of the cache is $\frac{M}{N}=\frac{Z}{F}$. In the delivery phase, the server makes a transmission corresponding to every integer present in the EPDA considered. Thus there are $S$ transmissions, each of size $\left(\frac{1}{F}\right)^{th}$ of a file. So, the normalized delivery time is $T=\frac{S}{F}$.
	
	Now, it remains to show the decodability of the demanded files by the users. Consider user $k$ which does not have the subfile $W_{d_k,j}$ from the placement phase. Assume that $a_{j,k} = s$ for some $s\in [S]$ (if $a_{j,k}=\star$, then $W_{n,j}$ would have been available for user $k$ from the placement phase itself). Then the claim is that user $k$ will receive the subfile $W_{d_k,j}$ from the transmission corresponding to the integer $s$. Consider the sub-array $\textbf{A}^{(s)}$, and assume that $a_{j,k}=a_{j_2,k_2}=\dots=a_{j_{g_s},k_{g_s}}=s$. Then the server transmission is 
	$\mathbf{V}^s.(W_{d_{k},j},W_{d_{k_2},j_2},\dots,W_{d_{k_{g_s}},j_{g_s}})^T$. But, user $k$ receives 
	\begin{equation*}
	Y_k = \mathbf{h}_k^T \mathbf{V}^s\cdot(W_{d_{k},j},W_{d_{k_2},j_2},\dots,W_{d_{k_{g_s}},j_{g_s}})^T+w_k
	\end{equation*}    
	Neglecting the noise component $w_k$ (using the high-SNR assumption),
	\begin{align*}
	Y_k &= \mathbf{h}_k^T \mathbf{V}^s\cdot(W_{d_{k},j},W_{d_{k_2},j_2},\dots,W_{d_{k_{g_s}},j_{g_s}})^T\\
	& = \mathbf{h}_k^T (\mathbf{v}^s_{1},\mathbf{v}^s_{2},\dots,\mathbf{v}^s_{{g_s}})\cdot(W_{d_{k},j},W_{d_{k_2},j_2},\dots,W_{d_{k_{g_s}},j_{g_s}})^T
	\end{align*}
	From the design of the precoding vectors, we have, $h_k^T\mathbf{v}_i^s = 0$ for all $i$ such that $a_{j,k}\neq\star$ and $a_{j,i}=s $, $j\in [F]$. In other words, the precoding vectors are designed such that whichever subfiles (subfiles involved in the transmission corresponding to the integer $s$) are not cached in the $k^{th}$ cache will be nulled out from $Y_k$ (except the subfile required for user $k$). Therefore, $Y_k$ will be a linear combination of required subfile (for user $k$) and some other subfiles that are available in cache $k$. All the users know all the channel coefficients $\textbf{H}$ completely, hence users $k$ can decode $W_{d_k,j}$. Since $k$ and $j$ are arbitrary, all the users can decode their demanded files. 
	
	This completes the proof of Theorem \ref{thm:ata}.
\end{IEEEproof}

\begin{exmp}
	\label{example1}
	Consider the $(K=4,L=2,F=4,Z=1,S=4)$ EPDA given below:
		\[
	\textbf{A} = \begin{bmatrix}
	\star & 1     & 1      & 4     \\
	1     & \star & 2      & 2     \\
	3     & 2     & \star  & 3     \\ 
	4     & 4     & 3      & \star
	\end{bmatrix}
	\]
	
	Consider a coded caching scheme with $K=4$ users and the server having $L=2$ transmit antennas. The server has $N$ files $W_n,n\in [N]$. The placement and delivery is in accordance with $\textbf{A}$. In the placement phase, the server divides each file into 4 subfiles, $W_n = \{W_{n,1},W_{n,2},W_{n,3},W_{n,4}\}$ for all $n\in [N]$. The contents stored in each user's cache are:
	\begin{align*}
		\mathcal{Z}_1 &= \left\{W_{n,1}:n\in[N]\right\},\\
		\mathcal{Z}_2 &= \left\{W_{n,2}:n\in[N]\right\},\\
		\mathcal{Z}_3 &= \left\{W_{n,3}:n\in[N]\right\},\\
		\mathcal{Z}_4 &= \left\{W_{n,4}:n\in[N]\right\}.
	\end{align*} 
	
	Let $\mathbf{d} = (1,2,3,4)$ be the demand vector. Then the server makes the transmissions summarized in TABLE \ref{table:exmp1}.

	\begin{table}[h!]
	\begin{center}
	\caption{Transmission in Example \ref{example1}}
	\label{table:exmp1}
	\begin{tabular}{cc}
	\hline \\ [-5pt]
	Time Slot &Transmission\\ [5pt]
	\hline \\ [-5pt]
	1 &  $\mathbf{V}^1\cdot(W_{1,2},W_{2,1},W_{3,1})^T$\\ [5pt]
	\hline \\ [-5pt]
	2 & $\mathbf{V}^2\cdot(W_{2,3},W_{3,2},W_{4,2})^T$\\ [5pt]
	\hline \\ [-5pt]
	3 & $\mathbf{V}^3\cdot(W_{1,3},W_{3,4},W_{4,3})^T$\\ [5pt]
	\hline \\ [-5pt]
	4 & $\mathbf{V}^4\cdot(W_{1,4},W_{2,4},W_{4,1})^T$\\ [5pt]
	\hline
	\end{tabular}
	\end{center}
	\end{table}
	
	To design the precoding matrix $\mathbf{V}^1 = (\mathbf{v}^1_1,\mathbf{v}^1_2,\mathbf{v}^1_3)$, consider the sub-array corresponding to the integer 1, 
	\[
	\textbf{A}^{(1)} = \begin{bmatrix}
	\star & 1     & 1 \\
	1     & \star & 2 
	\end{bmatrix}.
	\]
	To design the precoding vector $\mathbf{v}^1_1$, consider the row in which the integer 1 is present in the first column of $\textbf{A}^{(1)}$, that is the second row of $\textbf{A}^{(1)}$. Find out all the columns in which an integer is present in the second row of $\textbf{A}^{(1)}$ (excluding the first column). the integer 2 is present in the third column. Therefore, $\textbf{v}^1_1\perp \mathbf{h}_3$. Similarly, we can see that $\textbf{v}^1_2\perp \mathbf{h}_3$ and $\textbf{v}^1_3\perp \mathbf{h}_2$. By considering $\textbf{A}^{(2)}$, $\textbf{A}^{(3)}$ and $\textbf{A}^{(4)}$, we can design the rest of the precoding matrices.
	
	Now, we see how the users benefit from the transmission corresponding to integer 1. From the transmission corresponding to integer 1, user 1 (note that, $a_{2,1} = 1$) receives,
   \begin{align*}
	Y_1  &=\mathbf{h}_1^T\mathbf{V}^1\cdot(W_{1,2},W_{2,1},W_{3,1})^T\\
	     &=\mathbf{h}_1^T\cdot (\mathbf{v}^1_1,\mathbf{v}^1_2,\mathbf{v}^1_3)\cdot(W_{1,2},W_{2,1},W_{3,1})^T\\
	     &= (\mathbf{h}_1^T\mathbf{v}^1_1,\mathbf{h}_1^T\mathbf{v}^1_2,\mathbf{h}_1^T\mathbf{v}^1_3)\cdot(W_{1,2},W_{2,1},W_{3,1})^T
	\end{align*}
	Since, user 1 has access to the subfiles $W_{n,1}$, the user can null out the subfiles $W_{2,1}$ and $W_{3,1}$, and can get the desired subfile $W_{1,2}$. From the transmission corresponding to integer 1, user 2 (note that, $a_{1,2} = 1$) receives,
	   \begin{align*}
	Y_2  &=\mathbf{h}_2^T\mathbf{V}^1\cdot(W_{1,2},W_{2,1},W_{3,1})^T\\
	&=(\mathbf{h}_2^T\mathbf{v}^1_1,\mathbf{h}_2^T\mathbf{v}^1_2,\mathbf{h}_2^T\mathbf{v}^1_3)\cdot(W_{1,2},W_{2,1},W_{3,1})^T\\
	&= (\mathbf{h}_2^T\mathbf{v}^1_1,\mathbf{h}_2^T\mathbf{v}^1_2,0)\cdot(W_{1,2},W_{2,1},W_{3,1})^T\\
	&= \mathbf{h}_2^T\mathbf{v}^1_1W_{1,2}+\mathbf{h}_2^T\mathbf{v}^1_2W_{2,1}.
	\end{align*}
	Since, user 2 has access to the subfile $W_{1,2}$, the user can decode the desired subfile $W_{2,1}$. Finally, from the transmission corresponding to integer 1, user 3 (note that, $a_{1,3} = 1$) receives, 
		   \begin{align*}
	Y_3  &=\mathbf{h}_3^T\mathbf{V}^1\cdot(W_{1,2},W_{2,1},W_{3,1})^T\\
	&=(\mathbf{h}_3^T\mathbf{v}^1_1,\mathbf{h}_3^T\mathbf{v}^1_2,\mathbf{h}_3^T\mathbf{v}^1_3)\cdot(W_{1,2},W_{2,1},W_{3,1})^T\\
	&= (0,0,\mathbf{h}_3^T\mathbf{v}^1_3)\cdot(W_{1,2},W_{2,1},W_{3,1})^T\\
	&= \mathbf{h}_3^T\mathbf{v}^1_3W_{3,1}.
	\end{align*}
	That is, user 3 can get the desired subfile $W_{3,1}$. This completes the decodability of the subfiles involved in the transmission corresponding to integer 1. Similarly, there are transmissions corresponding to the remaining integers. From those transmissions, the users can decode the remaining subfiles needed. 
\end{exmp}

\begin{rem}
	For a given $K,L,F,Z$ and $S$, it is possible to have more than one EPDA. For example, the array $\textbf{A}_1$ and $\textbf{A}_2$ are $(K=3,L=2,F=3,Z=1,S=2)$ EPDAs.
	\[
	\textbf{A}_1 = \begin{bmatrix}
	\star & 2     & 1 \\
	1     & \star & 2 \\
	2     & 1     & \star 
	\end{bmatrix} \hspace{0.75cm}
\textbf{A}_2 = \begin{bmatrix}
	\star & 1     & 1 \\
	1     & \star & 2 \\
	2     & 2     & \star 
\end{bmatrix}
	\]
\end{rem}

\begin{rem}
	\label{rem:ata}
	If $L=1$, no row of $\textbf{A}^{(s)}$ should contain more than one integer. That is, every row in $\textbf{A}^{(s)}$ will be consisting of one integer and rest all $\star$'s. That one integer will be $s$, since $\textbf{A}^{(s)}$ is obtained by deleting all the rows and columns that do not contain the integer $s$. Therefore, $\textbf{A}^{(s)}$ will have the form, 
		\[
	 \begin{bmatrix}
	s      & \star   &\dots  & \star \\
	\star  & s       &\dots  & \star \\
	\vdots & \vdots  &\ddots  & \vdots \\
	\star  & \star   &\dots  & s
	\end{bmatrix}
	\]
	up to row and column permutation. That means, a $(K,L,F,Z,S)$ EPDA becomes a $(K,F,Z,S)$ PDA introduced in \cite{YCTC}, when $L=1$ (Conditions D1 and D2 for being a PDA are the same as the conditions C1 and C2 for being an EPDA).
\end{rem}

In a $(K,L,F,Z,S)$ EPDA, if $g_1=g_2=\dots,g_S =g$, then the EPDA is said to be $g$-regular.
\begin{defn}
	An array $\textbf{A}$ is said to be a $g$-regular $(K,L,F,Z,S)$ EPDA (or simply $g-(K,L,F,Z,S)$ EPDA) if $\textbf{A}$ satisfies the condition\textit{ C2'}, in addition to \textit{C1, C3} and \textit{C4}.\\ 
	C2': Each integer in $[S]$ should appear exactly $g$ times in $A$. 
\end{defn}

In a multi-antenna coded caching scheme obtained from a $g$-regular $(K,L,F,Z,S)$ EPDA, the positive integer $g$ represents the number of users benefited from a transmission in any given time slot. In other words, $g$ is the DoF achieved by the scheme. By \textit{C2}, the regularity of an EPDA $g$ is upper bounded by the number of columns $K$ in the EPDA.
\begin{lem}
	\label{lem:reg}
	For a multi-antenna coded caching scheme obtained from a $g$-regular $(K,L,F,Z,S)$ EPDA, the delivery time,\begin{equation}
		\label{reg:T}
		T=\frac{K}{g}\left(1-\frac{Z}{F}\right).
	\end{equation}
Furthermore, we have,
\begin{equation}
	\label{regineq}
	g\leq L+\frac{KZ}{F}.
\end{equation}
\end{lem}

\begin{IEEEproof}
	We prove Lemma \ref{lem:reg} using counting arguments. Consider a $g$-regular $(K,L,F,Z,S)$ EPDA. Let us count the total number of integers in the EPDA in two different ways. Each of the $S$ integers appear exactly $g$ times in the array. Similarly, there are $F-Z$ integers in each column of the EPDA. Therefore,
	\begin{equation*}
		gS=K(F-Z).
	\end{equation*}
	So, we have,
	\begin{align*}
		T = \frac{S}{F}=\frac{K}{g}\left(1-\frac{Z}{F}\right).
	\end{align*}

To prove \eqref{regineq}, we count the number of stars in the array in two different ways. Since, each column consists of $Z$ stars, total number of stars in the EPDA is $KZ$. Now, let $\delta_j$ be the number stars in the $j^{th}$ row of the EPDA. Consider an integer $s\in [S]$ present in the $j^{th}$ row. Then, in the corresponding row of $\textbf{A}^{(s)}$, number of integers is less than or equal to $L$. Therefore,
\begin{equation}
	\label{deltaj}
	\delta_j\geq g-L.
\end{equation} 
If no integer is present in the $j^{th}$ row, then the number of stars in that row is $K$. Even then \eqref{deltaj} is valid. Therefore we have,
\begin{equation}
	\sum_{j=1}^F\delta_j = KZ.
\end{equation}
Then,
\begin{equation*}
	F(g-L)\leq KZ.
\end{equation*}
Upon rearrangement, we have,
\begin{equation*}
		g\leq L+\frac{KZ}{F},
\end{equation*}
where the equality holds when $\delta_j=\frac{KZ}{F}$ for all $j\in [F]$. 

When $\frac{Z}{F}=\frac{t}{K}$ for some integer $t$, we have,
\begin{equation*}
	g\leq L+t.
\end{equation*}

This completes the proof of Lemma \ref{lem:reg}.
\end{IEEEproof}
\section{New Constructions}
\label{constructions}
In this section, we introduce two constructions of EPDAs for certain values of $K,L,F,Z,S$. Using those EPDAs, we obtain the corresponding multi-antenna coded caching schemes. Before dealing with the constructions, we present the definition of $u$-row concatenation of arrays. 
\begin{defn}
		The process of obtaining an array $\textbf{A}_{F \times uK}$ by concatenating another array $\hat{\textbf{A}}_{F\times K}$, row-wise, $u$ times is referred to as $u$-row concatenation of $\hat{\textbf{A}}$. This can be expressed as,
	\begin{equation*}
	\mathbf{A} = [\underbrace{\hat{\textbf{A}}|\hat{\textbf{A}}|\dots|\hat{\textbf{A}}}_{u \text{ times}}].
	\end{equation*}
\end{defn}

\begin{lem}
	\label{con:ATA}
	The $u$-row concatenation of a $(K,L,F,Z,S)$ EPDA results in a $(uK,uL,F,Z,S)$ EPDA.
\end{lem}
\begin{IEEEproof}
	Consider a $(K,L,F,Z,S)$ EPDA $\mathbf{A}_1$. Let $\textbf{A}$ be the array obtained by $u$-row concatenation of $\mathbf{A}_1$. That is,
	\begin{equation*}
	\mathbf{A} = [\underbrace{\mathbf{A}_1|\mathbf{A}_1|\dots|\mathbf{A}_1}_{u \text{ times}}].
	\end{equation*}
	Now, we prove Lemma \ref{con:ATA} by verifying that $\textbf{A}$ satisfies the conditions \textit{C1, C2, C3} and \textit{C4}.
	It is easy to see that the array $\textbf{A}$ satisfies \textit{C1, C2} and \textit{C3}, since $\mathbf{A}_1$ satisfies \textit{C1, C2} and \textit{C3}. Now to show $\textbf{A}$ satisfies \textit{C4}, consider an integer $s\in [S]$ present in $\textbf{A}_1$. The sub-array of $\textbf{A}$ corresponding to $s$ is obtained by the $u$-row concatenation of the sub-array of $\textbf{A}_1$ corresponding to $s$. That is,
	 \begin{equation*}
	 \mathbf{A}^{(s)} = [\underbrace{\mathbf{A}_1^{(s)}|\mathbf{A}_1^{(s)}|\dots|\mathbf{A}_1^{(s)}}_{u \text{ times}}].
	 \end{equation*}
	 In any row of $\mathbf{A}_1^{(s)}$, there will be $L$ integers at the maximum. It means that, no row of $\textbf{A}^{(s)}$ will be containing more than $uL$ integers. This is true for every $s\in [S]$. Therefore, $\textbf{A}$ is a $(uK,uL,F,Z,S)$ EPDA.
	 
	 This completes the proof of Lemma \ref{con:ATA}.
\end{IEEEproof}
\begin{exmp}
	Let $\textbf{A}_1$ be a $(3,2,3,1,2)$ EPDA.
		\[
	\textbf{A}_1 = \begin{bmatrix}
	\star & 1     & 1      \\
	1     & \star & 2       \\
	2    & 2     & \star  
	\end{bmatrix}
	\]
	
	The 2-row concatenation of $\textbf{A}_1$ results in the array $\textbf{A}$.
			\[
	\textbf{A} =[\textbf{A}_1|\textbf{A}_1] = \begin{bmatrix}
	\star & 1     & 1  &| &\star & 1     & 1    \\
	1     & \star & 2   &| &  1     & \star & 2 \\
	2    & 2     & \star &| &2    & 2     & \star
	\end{bmatrix}
	\]
	The array $\textbf{A}$ is a $(6,4,3,1,2)$ EPDA. 
\end{exmp}

\begin{corollary}
	\label{con:PDA}
	The $u$-row concatenation of a $(K,F,Z,S)$ PDA results in a $(uK,u,F,Z,S)$ EPDA.
\end{corollary}
\begin{IEEEproof}
	The result follows from Lemma \ref{con:ATA} since a $(K,F,Z,S)$ PDA is a $(K,L=1,F,Z,S)$ EPDA.
\end{IEEEproof}
\begin{exmp}
	A $(3,3,1,3)$ PDA $\textbf{P}_1$ is given below. 
	\[
	\textbf{P}_1 = \begin{bmatrix}
	\star & 1     & 3     \\
	1     & \star & 2       \\
3    & 2     & \star  
	\end{bmatrix}
	\]
	The array $\textbf{A}$ obtained by the 2-row concatenation of $\textbf{P}_1$ is a $(6,2,3,1,3)$ EPDA.
	\[
	\textbf{A} =[\textbf{P}_1|\textbf{P}_1] = \begin{bmatrix}
	\star & 1     & 3   &| &\star & 1     & 3   \\
	1     & \star & 2    &| &  1     & \star & 2 \\
	3    & 2     & \star  &| &3    & 2     & \star
	\end{bmatrix}
	\]
\end{exmp}
\begin{rem}
	The $u$-row concatenation of a $g$-regular $(K,L,F,Z,S)$ EPDA results in a $ug$-regular $(uK,uL,F,Z,S)$ EPDA. 
\end{rem}

\begin{rem}
\label{rem:LaE}	
	The multi-antenna coded caching scheme presented in \cite{LaE} can also be obtained from EPDA. The proposed scheme work with subpacketization number $\binom{K/L}{t/L}$ if $L|K$ and $L|t$, where $t$ is a positive integer such that $\frac{M}{N}=\frac{t}{K}$. To see the corresponding EPDA representation, first consider the $(K',F',Z'S')$ MaN PDA (given in \cite{YCTC}) with $K'=K/L, F'=\binom{K/L}{t/L},Z'=\binom{(K/L)-1}{(t/L)-1}$ and $S'=\binom{K/L}{(t/L)+1}$. The $L$-row concatenation of this $(K',F',Z',S')$ PDA will give our desired $(K,L,F',Z'S')$ EPDA (follows from Corollary \ref{con:PDA}). Note that, in the resulting $(K,L,M,N)$ multi-antenna coded caching scheme, the subpacketization number is $\binom{K/L}{t/L}$, and the delivery time is $\frac{S'}{F'}=\frac{K-t}{t+L}$ (DoF achieved is $t+L$). 
	
\end{rem}

Now, we see two constructions of EPDAs with the maximum possible regularity $L+\frac{KZ}{F}$. 
\begin{itemize}
	\item \textbf{Construction \Romannum{1}}: A $(K,K-Z,K,Z,K-Z)$ EPDA. 
	\item \textbf{Construction \Romannum{2}}: A $\left(K,\frac{K-nZ}{n-1},K,Z,(n-1)K\right)$ EPDA with $K\geq (2n-1)Z$, where $n$ is a positive integer greater than 1.
\end{itemize}

 First, we present Construction \Romannum{1}.
 
a) \textbf{Construction \Romannum{1}}

In this construction, $F=K$, $L=K-Z$ and $S=K-Z$. We denote the EPDA with $\textbf{A}=[a_{j,k}]$. Then, the $\star$'s appear in $\textbf{A}$ as follows,
\begin{equation}
a_{j,k} = \star, \hspace{0.3cm} \forall (j,k)\in [K]\times[K]: (j-k) \hspace{0.1cm}(mod\hspace{0.1cm}K)< Z.
\end{equation}
That is, in the $k^{th}$ column of $\textbf{A}$, $a_{k,k}=a_{<k+1>_K,k}=\dots=a_{<k+Z-1>_K,k}=\star$. Therefore, there will be exactly $Z$ $\star$'s in each column. Now, consider an integer $s\in [K-Z]$. Then $s$ occurs in $\textbf{A}$ as,
\begin{equation}
a_{j,k} = s, \text{ such that } j= (Z+s+k-1) \hspace{0.05cm}(mod\hspace{0.05cm}K). 
\end{equation}

Now we have to verify that the array $\textbf{A}$ obtained using Construction \Romannum{1} is, in fact, an EPDA. We have already seen that, $\star$ appears $Z$ times in each column (\textit{C1} is satisfied). Also, every integer in the set $[K-Z]$ appears $K$ times in $\textbf{A}$ (\textit{C2'} is also satisfied). In the $k^{th}$ column of $\textbf{A}$, the integer $s\in[K-Z]$ appears only in the $(Z+s+k-1) \hspace{0.05cm}(mod\hspace{0.05cm}K)$-th row. Since no integer appears more than once in any column, the array $\textbf{A}$ satisfies \textit{C3} as well. As we have seen, every integer will be present in all the columns. Also, since the set $\{(Z+s+k-1\}_k^K=[K]$ for a given $s$, the integer $s$ will appear in all the rows as well. Therefore, for every $s\in [S]$,
\begin{equation*}
\textbf{A}^{(s)} = \textbf{A}.
\end{equation*} 
By the construction, every row consists of exactly $Z$ $\star$'s. So, number of integers in any row is $K-Z$. Since $L=K-Z$, the array $\textbf{A}$ satisfies \textit{C4}. Hence the array $\textbf{A}$ obtained using the Construction \Romannum{1} is a $K$-regular $(K,K-Z,K,Z,K-Z)$ EPDA. 
\begin{exmp}
	We present an example for a $4$-regular $(4,3,4,1,3)$ EPDA obtained using Construction \Romannum{1}.
	\[
	\textbf{A} = \begin{bmatrix}
	\star & 3     & 2      & 1     \\
	1     & \star & 3      & 2     \\
	2     & 1     & \star  & 3     \\ 
	3     & 2     & 1      & \star
	\end{bmatrix}
	\]
	Note that, the sub-arrays corresponding to the integers $\textbf{A}^{(1)}=\textbf{A}^{(2)}=\textbf{A}^{(3)}=\textbf{A}$.
\end{exmp}

\begin{thm}
	\label{thm:constA}
	For a $(K,L,M,N)$ multi-antenna coded caching scheme with $\frac{M}{N}=\frac{t}{K}$ and $L=K-t$, where $t$ is an integer, the delivery time $T^*=\frac{K-t}{t+L}=\frac{L}{K}$ is achievable with a subpacketization number $\frac{K}{\text{gcd}(K,t,L)}$. 
\end{thm}
\begin{IEEEproof}
	Let $t=\frac{KM}{N}$ be an integer, and let $L=K-t$. Define $\gamma \triangleq \text{gcd}(K,t,L)$. Let $\tilde{K}=\frac{K}{\gamma},\tilde{L}=\frac{L}{\gamma}$ and $\tilde{t}=\frac{t}{\gamma}$. Since $K=t+L$, we have, $\tilde{K}=\tilde{t}+\tilde{L}$. Now construct a $(\tilde{K},\tilde{L},\tilde{K},\tilde{t},\tilde{L})$ EPDA $\textbf{A}_1$ using Construction \Romannum{1}. The $\gamma$-row concatenation of $\textbf{A}_1$ results in a $(K,L,\tilde{K},\tilde{t},\tilde{L})$ EPDA $\textbf{A}$. Now, in the multi-antenna coded caching scheme corresponding to $\textbf{A}$, $\frac{M}{N} = \frac{\tilde{t}}{\tilde{K}}=\frac{t}{K}$. Using Theorem \ref{thm:ata}, in the resulting scheme, we have, the delivery time,
	\begin{equation}
	T=\frac{\tilde{L}}{\tilde{K}}=\frac{K-t}{t+L},
	\end{equation} 
	and the subpacketization number is $\frac{K}{\gamma}$. This completes the proof of Theorem \ref{thm:constA}.
\end{IEEEproof}

Now, we present Construction \Romannum{2}.

b) \textbf{Construction \Romannum{2}}

 We construct an EPDA $\textbf{B}=[b_{j,k}]$ with parameters $F=K, L=\frac{K-nZ}{n-1}$,  $S=(n-1)K$ and $K\geq (2n-1)Z$, where $n$ is a positive integer greater than 1. The $\star$'s appear in $\textbf{B}$ as follows,
 \begin{equation}
 \label{star:B}
 b_{j,k} = \star, \hspace{0.3cm} \forall (j,k)\in [K]\times[K]: (j-k) \hspace{0.1cm}(mod\hspace{0.1cm}K)< Z.
 \end{equation}
 That is, in the $k^{th}$ column of $\textbf{B}$, $b_{k,k}=b_{<k+1>_K,k}=\dots=b_{<k+Z-1>_K,k}=\star$. Therefore, there will be exactly $Z$ $\star$'s in each column. Now, we see the appearance of integers in the array $\textbf{B}$. An integer will be appearing $Z+L$ times in the array. One integer will be appearing in two rows: $Z$ times in one row and $L$ times in the other. Consider an integer $s\in [(n-1)K]$. Let $s =pK+q$, where $p\in [0:n-2], q\in [1:K]$.  If $p$ is even, then $s$ occurs in $\textbf{B}$ as follows:
 \begin{align}
 b_{q,<\frac{p}{2}(Z+L)+q+i>_K} &= s \hspace{0.3cm} \forall i \in [L],\label{int:Beven1}\\
 b_{<\frac{p}{2}(Z+L)+Z+q>_K,<q-Z+i>_K} &= s \hspace{0.3cm} \forall i \in [Z].\label{int:Beven2}
 \end{align}
 If $p$ is odd, then $s$ occurs in $\textbf{B}$ as follows:
 \begin{align}
b_{q,<(\frac{p-1}{2})(Z+L)+L+q+i>_K} &= s \hspace{0.3cm} \forall i \in [Z],\label{int:Bodd1}\\
b_{<(\frac{p+1}{2})(Z+L)+q>_K,<q-Z+i>_K} &= s \hspace{0.3cm} \forall i \in [L].\label{int:Bodd2}
\end{align}
The $\star$'s occur as the diagonal entries and in the $Z-1$ rows below the diagonal entries (in every column) in $\textbf{B}$. When $p=0$ ($s\in[K]$), from \eqref{int:Beven1}, \eqref{int:Beven2}, the integers will occur adjacent to the $\star's$ off diagonally (above and below the diagonal). As $p$ increases ($p\in [1:n-2]$), the integers occur away from the diagonal (for a fixed $p$, the integers for different $q\in [K]$ appear in $K$ different rows, but at positions that are away from the diagonal by the same amount). 

Now, it remains to verify that the array $\textbf{B}$ obtained from Construction \Romannum{2} is, in fact, an EPDA. From \eqref{star:B}, it is clear that $\textbf{B}$ satisfies \textit{C1}. Every integer in the set $[(n-1)K]$ appears exactly $Z+L$ times in the array. Therefore, $\textbf{B}$ satisfies \textit{C2'} as well. To show that the array $\textbf{B}$ satisfies \textit{C3}, consider the following two cases:\\
\textit{Case a)} When $p$ is even, the integer $s=pK+q$ appear in the columns $<\frac{p}{2}(Z+L)+q+\alpha>_K$ for $\alpha \in [L]$ and $<q-Z+\beta>_K$ for $\beta \in [Z]$. But, 
$$<\frac{p}{2}(Z+L)+q+\alpha>_K \hspace{0.1cm}\in \hspace{0.1cm}[<q+1>_K:<q+\frac{K-L}{2}>_K]_K,$$ 
where, 
$$<q-Z+\beta>_K \hspace{0.1cm}\in \hspace{0.1cm}[<q-Z>_K:q]_K.$$ 
The sets $[<q+1>_K:<q+\frac{K-L}{2}>_K]_K$ and $[<q-Z>_K:q]_K$ are non-overlapping since $K$ is at least $2Z+L$. \\
\textit{Case b)} When $p$ is odd, the integer $s=pK+q$ appear in the columns $<(\frac{p-1}{2})(Z+L)+L+q+\alpha>_K$ for $\alpha \in [Z]$ and $<q-Z+\beta>_K$ for $\beta \in [L]$. But, 
\begin{align*}
<(\frac{p-1}{2})(Z+L)+&L+q+\alpha>_K\hspace{0.1cm}\in\\ &\hspace{0.1cm}[<q+L+1>_K:<q+\frac{K-Z}{2}>]_K,
\end{align*}
where,
$$<q-Z+\beta>_K \hspace{0.1cm}\in \hspace{0.1cm}[<q-Z>_K:<q+L-Z>]_K.$$
Since the sets $[<q+L+1>_K:<q+\frac{K-Z}{2}>]_K$ and $[<q-Z>_K:<q+L-Z>]_K$ are non-overlapping, the integer $s$ will appear at most once in a column. Therefore, the array satisfies\textit{ C3} as well. 

To show that $\textbf{B}$ satisfies \textit{C4}, again consider two cases:\\
\textit{Case a)} When $p$ is even, the integer $s=pK+q$ appears $L$ times in the $q^{th}$ row and $Z$ times in the $<\frac{p}{2}(Z+L)+Z+q>_K$-$th$ row. In the $<\frac{p}{2}(Z+L)+Z+q>_K$-$th$ row,  $b_{<\frac{p}{2}(Z+L)+Z+q>_K,q}=b_{<\frac{p}{2}(Z+L)+Z+q>_K,<q-1>_K}=\dots=b_{<\frac{p}{2}(Z+L)+Z+q>_K,<q-Z+1>_K}\text{\hspace{-0.09cm}}=\text{\hspace{-0.06cm}}s$. At the same time in the $q^{th}$ row, $b_{q,q}=b_{q,<q-1>_K}=\dots=b_{q,<q-Z+1>_K}=\star$.  Similarly, in the $q^{th}$ row, $b_{q,j}=s$ for all $j\in [<\frac{p}{2}(Z+L)+q+1>_K:<\frac{p}{2}(Z+L)+q+Z>_K]_K$ if $L\geq Z$. But, in the $<\frac{p}{2}(Z+L)+Z+q>_K-th$ row, $b_{<\frac{p}{2}(Z+L)+Z+q>_K,j}=\star$ for all $j\in [<\frac{p}{2}(Z+L)+q+1>_K:<\frac{p}{2}(Z+L)+q+Z>_K]_K$. Therefore, both the rows in the sub-array $\textbf{B}^{(s)}$ consist of at most $L$ integers, if $L\geq Z$. The condition $L\geq Z$ ensured by the fact that $K\geq (2n-1)Z$. \\
\textit{Case b) }When $p$ is odd, the integer $s=pK+q$ appears $Z$ times in the $q^{th}$ row and $L$ times in the $<(\frac{p+1}{2})(Z+L)+q>_K$-$th$ row. In the $<(\frac{p+1}{2})(Z+L)+q>_K$-$th$ row,  $b_{<(\frac{p+1}{2})(Z+L)+q>_K,j}=s$, for all $j\in [<q-Z+1>_K:q]_K$, if $L\geq Z$. But, in the $q^{th}$ row, $b_{q,j}=\star$, for all $j\in [<q-Z+1>_K:q]_K$. Similarly, in the $q^{th}$ row $b_{q,j}=s$, for all $j\in [<(\frac{p-1}{2})(Z+L)+L+q>_K:<(\frac{p+1}{2})(Z+L)+q>_K]_K$. At the same time, in the $<(\frac{p+1}{2})(Z+L)+q>_K$-$th$ row, $b_{<(\frac{p+1}{2})(Z+L)+q>_K,j}=\star$, for all $j\in [<(\frac{p-1}{2})(Z+L)+L+q>_K:<(\frac{p+1}{2})(Z+L)+q>_K]_K$. Therefore, both the rows in the sub-array $\textbf{B}^{(s)}$ consist of at most $L$ integers.\\
In summary, for every $s\in [(n-1)K]$, the rows in the sub-array $\textbf{B}^{(s)}$ consist of at the most $L$ integers. In other words, the array $\textbf{B}$ satisfies \textit{C4}. It means that, $\textbf{B}$ is a $(Z+L)$-regular $(K,L,K,Z,(n-1)K)$ EPDA with $L=\frac{K-nZ}{n-1}$ and $K\geq (2n-1)Z$. 

\begin{exmp}
	We see an example for a $3$-regular $(4,2,4,1,4)$ EPDA obtained using Construction \Romannum{2}. We have, $K=nZ+(n-1)L$ where $n=2$. 
	
	\[
	\textbf{B} = \begin{bmatrix}
	\star & 1     & 1      & 4     \\
	1     & \star & 2      & 2     \\
	3     & 2     & \star  & 3     \\ 
	4     & 4     & 3      & \star
	\end{bmatrix}
	\]
	All the four integers are appearing thrice in the array. The sub-array corresponding to the integer 1 is
		\[
	\textbf{B}^{(1)} = \begin{bmatrix}
	\star & 1     & 1    \\
	1     & \star & 2   
	\end{bmatrix}.
	\]
	The integer 1 appears $L=2$ times in the first row and $Z=1$ time in the second row. The same is the case with all the remaining three integers.
\end{exmp}

Now, we see a bigger example of an EPDA constructed using Construction \Romannum{2}.

\begin{exmp}
	A $(17,3,17,2,51)$ EPDA with regularity $Z+L=5$ obtained using Construction \Romannum{2} is given in \eqref{eqn:matrix}. In this example, we have, $K=nZ+(n-1)L$ with $n=4$.
	\begin{figure*}[t]
		\begin{equation}
		\label{eqn:matrix}	
		\begin{aligned}	
		\textbf{B} = \begin{bmatrix}
		\star & 1     &  1& 1     & 18& 18&    35&   35&  35& 45&  45& 30& 30& 30& 16& 16&\star  \\
		\star & \star &  2& 2     &  2& 19&   19&   36&  36& 36&  46&  46& 31& 31& 31& 17& 17 \\
		1 & \star & \star & 3     &  3&  3&   20&   20&  37& 37& 37& 47& 47& 32& 32& 32& 1 \\
		2 & 2     & \star & \star &  4&  4&    4&  21& 21& 38& 38& 38& 48& 48& 33& 33& 33 \\
		34 & 3     & 3    & \star & \star &  5& 5& 	5& 22& 22& 39& 39& 39& 49& 49& 34& 34 \\
		18 & 18    & 4    & 4     & \star & \star & 6&  6&  6&23&23& 40& 40& 40& 50& 50& 18 \\
		19 & 19    & 19   & 5     &  5& \star &\star &	7&  7& 7&24&24& 41& 41& 41& 51& 51 \\
		35 & 20    & 20    & 20    &  6&  6&\star &\star & 8& 	8& 8&25&25& 42& 42& 42& 35 \\
		36 & 36    & 21    & 21    &     21& 7& 	7&\star &\star & 9& 9& 9&26&26& 43& 43& 43 \\
		44 & 37    &37     & 22    &     22&     22&  8&  8&\star &\star & 10& 10& 10&27&27& 44& 44 \\
		45 & 45    & 38    & 38    &     23&     23& 23&  9& 9&\star &\star & 11& 11& 11&28&28& 45 \\
		46 & 46    & 46    & 39    &     39&     24& 24& 24&10& 10&\star &\star & 12& 12& 12&29& 29 \\
		30& 47    & 47    & 47    &     40&     40& 25& 25&25& 11& 11&\star &\star & 13& 13& 13& 30 \\
		31 & 31     & 48   &48     &     48&     41& 41& 26&26&26& 12& 12&\star &\star & 14& 14& 14 \\
		15 & 32    & 32   & 49     &     49&     49& 42& 42&27&27&27& 13& 13&\star &\star & 15& 15 \\
		16 & 16   & 33    & 33    &      50&      50&  50&  43& 43&28&28&28& 14& 14&\star &\star &16 \\
		17 & 17   & 17    & 34    &     34&      51&  51&  51& 44& 44&29&29&29& 15& 15&\star &\star  
		\end{bmatrix}	
		\end{aligned}
		\end{equation}
	\end{figure*}
\end{exmp}
\begin{thm}
	\label{thm:constB}
	For a $(K,L,M,N)$ multi-antenna coded caching scheme with $\frac{M}{N}=\frac{t}{K}$ and $K=nt+(n-1)L; \hspace{0.1cm}L\geq t$, where $t$ and $n\geq 2$ are integers, the delivery time $T^*=\frac{K-t}{t+L}=n-1$ is achievable with a subpacketization number $\frac{K}{\text{gcd}(K,t,L)}$.
\end{thm}
\begin{IEEEproof}
Let $t=\frac{KM}{N}$ be an integer, and let $n\geq 2$ be an integer such that $K=nt+(n-1)L$. Define $\gamma \triangleq \text{gcd}(K,t,L)$. Let $\tilde{K}=\frac{K}{\gamma},\tilde{L}=\frac{L}{\gamma}$ and $\tilde{t}=\frac{t}{\gamma}$. Since $K=nt+(n-1)L$, we have, $\tilde{K}=n\tilde{t}+(n-1)\tilde{L}$. Now construct a $(\tilde{K},\frac{\tilde{K}-n\tilde{t}}{n-1},\tilde{K},\tilde{t},(n-1)\tilde{K})$ EPDA $\textbf{B}_1$ using Construction \Romannum{2}. The $\gamma$-row concatenation of $\textbf{B}_1$ results in a $(K,\frac{K-nt}{n-1},\tilde{K},\tilde{t},\tilde{L})$ EPDA $\textbf{B}$. Now, in the multi-antenna coded caching scheme corresponding to $\textbf{B}$, $\frac{M}{N} = \frac{\tilde{t}}{\tilde{K}}=\frac{t}{K}$. Using Theorem \ref{thm:ata}, the delivery time obtained in the resulting scheme,
\begin{equation}
T=\frac{(n-1)\tilde{K}}{\tilde{K}}=n-1=\frac{K-t}{t+L},
\end{equation} 
and the subpacketization number is $\frac{K}{\gamma}$. This completes the proof of Theorem \ref{thm:constB}.	
\end{IEEEproof}
\section{Comparison}
\label{comparison}
In this section, we compare the multi-antenna coded caching schemes resulting from Construction \Romannum{1} and Construction \Romannum{2} to different multi-antenna coded caching schemes in the literature \cite{SMK,LaE,STS,SPSET}. All these schemes achieve the optimal DoF of $t+L$ (in the parameter settings for which the schemes are valid). Therefore, the comparison is in terms of subpacketization number. 

Consider the multi-antenna coded caching scheme obtained from Construction \Romannum{1}. The scheme requires a file to be split into $\frac{K}{\text{gcd}(K,t,L)}$ subfiles. When $K=t+L$, the multi-antenna scheme proposed in \cite{SMK} has a subpacketization number 
\begin{equation*}
	\binom{K}{t}\binom{K-t-1}{L-1} = \binom{K}{t}.
\end{equation*}
For any $t\in[2:K-2]$, the subpacketization number for our proposed scheme is strictly less than that for the scheme proposed in \cite{SMK}. The scheme in \cite{LaE} is applicable only when $L|K$ and $t|K$. When both those conditions are valid, along with $K=t+L$, then $t=iL$ and $K = (i+1)L$ for some positive integer $i$. In that case, the subpacketization number required for the scheme from Construction \Romannum{1} and for the scheme in \cite{LaE} is $\frac{K}{L}$ (because $\text{gcd}(K,t,L)=L$). The multi-antenna coded caching schemes in \cite{STS} and \cite{SPSET} are applicable when $L\geq t $. If $K=t+L$, then the subpacketization number required for the scheme in \cite{STS} is,
\begin{equation*}
	K(t+L) = K^2,
\end{equation*}
and that required for the scheme in \cite{SPSET} is,
\begin{equation*}
	\frac{K(t+L)}{\left(\text{gcd}(K,t,L)\right)^2} = \left(\frac{K}{\text{gcd}(K,t,L)}\right)^2.
\end{equation*}
Both these subpacketization requirements are high compared to our proposed scheme. The comparison between our scheme from Construction \Romannum{1} and the schemes in \cite{SMK,LaE,STS,SPSET} is summarized in Table \ref{table:constA}.
\begin{table}
	\caption{Comparison of subpacketization requirement for different multi-antenna coded caching schemes ($K=t+L$)}
	\label{table:constA}
	\renewcommand{\arraystretch}{1.8}
	\begin{center}
	\begin{tabular}{ |c|c| }
		\hline
		 Multi-antenna & Subpacketization Number\\ coded caching scheme & (when $K=t+L$) \\
		\hline
		Scheme from Construction \Romannum{1} & $\frac{K}{\gamma};\hspace{0.1cm}\gamma= \text{gcd}(K,t,L)$  \\
		\hline
	    Scheme in \cite{SMK} & $\binom{K}{t}$  \\
	    \hline
	     Scheme in \cite{LaE} & $\frac{K}{L}$ \\&(valid when $L|K$ and $L|t$) \\
	    \hline
	     Scheme in \cite{STS} & $K^2$ \\&(valid when $L\geq t$) \\
	    \hline
	    Scheme in \cite{SPSET} & $\left(\frac{K}{\gamma}\right)^2;\hspace{0.1cm}\gamma= \text{gcd}(K,t,L)$ \\&(valid when $L\geq t$) \\
	    \hline
	\end{tabular}
\end{center}
\renewcommand{\arraystretch}{1}
\end{table}

Now, we compare the multi-antenna coded caching scheme obtained from Construction \Romannum{2} with the schemes in \cite{SMK,LaE,STS,SPSET} in terms of subpacketization number. The comparison is for the case $K=nt+(n-1)L;\hspace{0.1cm}L\geq t$, where $n\geq2$ is a positive integer. The scheme from Construction \Romannum{2}  has a subpacketization number $\frac{K}{\text{gcd}(K,t,L)}$. When $K=nt+(n-1)L$, the multi-antenna scheme proposed in \cite{SMK} has a subpacketization number of
\begin{equation*}
\binom{K}{t}\binom{K-t-1}{L-1} = \binom{K}{t}\binom{(n-1)L-1}{L-1}.
\end{equation*}
The subpacketization number is increasing exponentially in the case of \cite{SMK}. Clearly, the scheme in \cite{SMK} has higher subpacketization number compared to our proposed scheme. Meaningful comparison with the scheme in \cite{LaE} is when $L=t$, since our proposed scheme works if $L\geq t$, and the scheme in \cite{LaE} works if $L|t$. When $t=L$, and $K=nt+(n-1)L$, our proposed scheme and the scheme in \cite{LaE} require a subpacketization number $\frac{K}{L}$ (because $\text{gcd}(K,t,L)=L$). The multi-antenna coded caching scheme in \cite{STS} requires a subpacketization number of $K(t+L)$ which is strictly greater than the subpacketization number required for our proposed scheme. Similarly, the scheme in \cite{SPSET} requires a subpacketization number
\begin{equation*}
\frac{K(t+L)}{\left(\text{gcd}(K,t,L)\right)^2} >\frac{K}{\text{gcd}(K,t,L)},
\end{equation*}
since $t+L>\text{gcd}(K,t,L)$. The comparison between our scheme from Construction \Romannum{2} and the schemes in \cite{SMK,LaE,STS,SPSET} is summarized in Table \ref{table:constB}.
\begin{table}
	\caption{Comparison of subpacketization requirement for different multi-antenna coded caching schemes ($K=nt+(n-1)L;L\geq t$)}
	\label{table:constB}
	\renewcommand{\arraystretch}{1.85}
	\begin{center}
		\begin{tabular}{ |c|c| } 
			\hline
			Multi-antenna& Subpacketization Number\\coded caching scheme & (when $K=nt+(n-1)L;\hspace{0.1cm}L\geq t$) \\
			\hline
			Scheme from Construction \Romannum{2} & $\frac{K}{\gamma};\hspace{0.1cm}\gamma= \text{gcd}(K,t,L)$  \\
			\hline
			Scheme in \cite{SMK} & $\binom{K}{t}\binom{(n-1)L-1}{L-1}$  \\
			\hline
			Scheme in \cite{LaE} & $\frac{K}{L}$ \\&(valid when $L=t$) \\
			\hline
			Scheme in \cite{STS} & $K(t+L)$ \\
			\hline
			Scheme in \cite{SPSET} & $\frac{K(t+L)}{\gamma^2};\hspace{0.1cm}\gamma= \text{gcd}(K,t,L)$ \\
			\hline
		\end{tabular}
	\end{center}
	\renewcommand{\arraystretch}{1}
\end{table}
\section{Conclusion}
\label{conclusion}
In this work, we proposed the idea of EPDAs to obtain $(K,L,M,N)$ multi-antenna coded caching schemes. The PDAs (proposed in \cite{YCTC} for the single-antenna centralized coded caching scheme) is a special class of EPDAs with $L=1$. We provided two constructions of EPDAs that result in multi-antenna coded caching schemes with very low subpacketization number. Also, we provided EPDA representation for the multi-antenna coded caching scheme proposed in \cite{LaE}.

In this work, we mainly focused on high SNR analysis, often neglecting the additive noise. It will be interesting to analyse the performance of the schemes in low and moderate SNR regimes. Also, observing the performance of the schemes resulting from EPDAs using different beamformers/precoders is a future direction to work on. Furthermore, we have seen that for the same parameter, setting there may exist more than one EPDAs. It will be exciting to compare the schemes obtained from those EPDAs. Above all, designing EPDAs for different values of $K,L,F,Z,S$ is of at most significance.

\section*{Acknowledgment}
This work was supported partly by the Science and Engineering Research Board (SERB) of Department of Science and Technology (DST), Government of India, through J.C. Bose National Fellowship to B. Sundar Rajan.

\end{document}